\documentclass[11pt]{article}
\usepackage{moriond,epsfig}

\def\Journal#1#2#3#4{{#1} {\bf #2}, #3 (#4)}

\def\NPB{{\em Nucl. Phys.} B}

\def\PRL{\em Phys. Rev. Lett.}
\def\PRD{{\em Phys. Rev.} D}

\def\be{\begin{equation}}
\def\ee{\end{equation}}
\def\bea{\begin{eqnarray}}
\def\eea{\end{eqnarray}}

\def\bzb{{\overline{B}^0_d}}
\def\bz{{B^0_d}}

\def\bplus{{B^+}}
\def\bzd{B^0_d}
\def\bzdb{\overline{B}^0_d}

\def\pip{\pi^+}
\def\piz{\pi^0}
\def\rhop{\rho^+}

\def\dz{D^0}

\def\dstm{D^{*-}}
\def\dstz{D^{*0}}

\def\ks{K^0_S}
\def\kl{K^0_L}

\def\fcp{f_{CP}}
\def\af{A_f}
\def\afb{\overline{A}_f}

\def\dmd{\Delta m_d}
\def\delt{\Delta t}
\def\dtp{\Delta t'}
\def\taubz{\tau(\bz)}
\def\taubm{\tau(\bplus)}
\def\taubp{\tau(\bplus)}
\def\lmbg{\lambda_{BG}}
\def\flmbg{f_{\lmbg}}

\def\micron{$\mu$m}

\begin{document}
\begin{flushright}
Belle-preprint 2001-8
\end{flushright}
\vspace*{-1cm}
\begin{flushleft}
\epsfxsize 1.4 truein \epsfbox{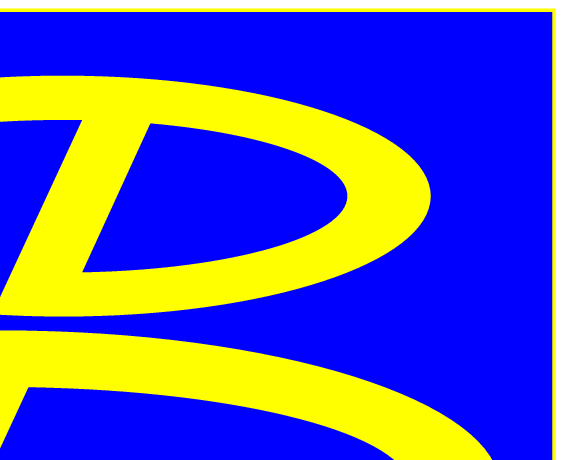}
\end{flushleft}

\vspace*{3cm}
\title{$B$ $CP$-VIOLATION, MIXING AND LIFETIME RESULTS FROM BELLE}

\author{ H. TAJIMA (Belle Collaboration)}

\address{Department of Physics, University of Tokyo,\\
7-3-1 Hongo, Bunkyo-ku, Tokyo 113-0033 Japan}

\maketitle

\bigskip
\bigskip
\bigskip

\abstracts{Recent results on the $B$ $CP$-violation
parameter $\sin2\phi_1$, the $\bzd$-$\bzdb$ mixing parameter $\dmd$ and $B$-meson lifetimes are reported.
The results are based on 10.5 $fb^{-1}$ of data collected with the Belle detector at KEKB.
Using both semileptonic and hadronic decay modes the neutral and charged $B$-meson lifetimes are measured to be $\taubz=1.548\pm0.035\;(\mathrm{stat.})$~ps and $\taubm=1.656\pm0.038\;(\mathrm{stat.})$~ps.
The oscillation frequency $\dmd$ for $\bzd$-$\bzdb$ mixing is measured to be $0.522\pm0.026\;(\mathrm{stat.})$~ps$^{-1}$ using semileptonic modes and $0.527\pm0.032\;(\mathrm{stat.})$~ps$^{-1}$ using hadronic modes.
The above results are preliminary.
The time evolution of $\bz$ decays into $J/\psi\ks$, $\psi(2S)\ks$, $\chi_{c1}\ks$, $\eta_{c}\ks$, $J/\psi\kl$ and  $J/\psi \pi^0$ modes are studied to obtain $\sin2\phi_1$.
Using 282 fully reconstructed events, $\sin2\phi_1$ is measured to be
$$\sin 2\phi_1=0.58^{+0.32}_{-0.34}\;(\mathrm{stat.})^{+0.09}_{-0.10}\;(\mathrm{syst.}).$$
}

\bigskip
\bigskip

\begin{center}
Contributed to the Proceedings of the  XXXVIth Rencontres de Moriond session devoted to QCD AND HIGH ENERGY HADRONIC INTERACTIONS, \\
March 17 --  24, 2001, Bourg-Saint-Maurice, France.
\end{center}

\newpage
$CP$ violation is one possible ingredience to understanding
 the antimatter deficit in the universe.
In the Kobayashi-Maskawa theory\cite{KM} $CP$ violation is due
to the complex phase of the quark mixing matrix.
According to this framework, a large asymmetry, ${\cal O}(1)$ can be expected in $B$ decays.\cite{carter}
However, the mechanism of the $CP$ violation has not yet been confirmed experimentally.
It is crucial to establish the $CP$ violation in $B$ decays expected from the KM theory.

$\bz\to J/\psi\ks$ is the most promising decay mode for observing
$CP$ violation in $B$ decays because of its low experimental
backgrounds and negligible theoretical uncertainties.
In the decays into $CP$ eigenstates ($\fcp$),
$CP$ violation occurs through interference of two decay amplitudes,
$\af e^{{-\Gamma t}/{2}}e^{{-iMt}/{2}}\cos(\dmd t)$
for $\bz\to\fcp$ and $\frac{q}{p}\afb e^{{-\Gamma t}/{2}}e^{{-iMt}/{2}}i\sin(\dmd t)$
for $\bz\to\bzb\to\fcp$, where $\Gamma$ is the total $\bz$ decay width, $\dmd$ the mass difference between the two $\bz$ mass eigenstates, $\af=\langle \fcp|{\cal H}|\bz \rangle$ and $\afb=\langle \fcp|{\cal H}|\bzb \rangle$.
Assuming negligible $CP$ violation in state mixing ($|{q}/{p}|\approx 1$) and direct $CP$ violation ($|\af/\afb|\approx 1$), the decay rates for $\bz\to\fcp$ and and its charge conjugate mode $\bzb\to\fcp$ can be expressed as
\begin{eqnarray}
\Gamma_{\bz\to\fcp}(t)&\approx&\af^2e^{-\Gamma t}\{1-{\cal I}m(\lambda)\sin(\dmd t)\}=\af^2e^{-\Gamma t}\{1+\xi_f\sin2\phi_1\sin(\dmd t)\},\nonumber \\
\Gamma_{\bzb\to\fcp}(t)&\approx&\af^2e^{-\Gamma t}\{1-{\cal I}m(1/\lambda)\sin(\dmd t)\}=\af^2e^{-\Gamma t}\{1-\xi_f\sin2\phi_1\sin(\dmd t)\}.
\end{eqnarray}
where $\lambda\equiv{q}/{p}\cdot{\af}/{\afb}$, $\xi_f$ is the CP eigenvalue of the $\fcp$ state, and $\phi_1$ is defined as $\phi_1\equiv \pi-\arg\left(\frac{-V^*_{tb}V_{td}}{-V^*_{cb}V_{cd}}\right)$.~\cite{Sanda}
$\bzd$-$\bzdb$ mixing gives ${q}/{p}\propto (V_{tb}^*V_{td})/(V_{tb}V_{td}^*)$.
When $\fcp$ is $(c\overline{c})K^0$ ($K^0\to \ks$ or $\kl$), $\af/\afb=\xi_{f}(V_{cb}^*V_{cs})/(V_{cb}V_{cs}^*)\cdot(V_{cs}^*V_{cd})/(V_{cs}V_{cd}^*)=\xi_f(V_{cb}^*V_{cd})/(V_{cb}V_{cd}^*)$.

In an asymmetric B factory, $\bz\bzb$ pairs are produced at the $\Upsilon(4S)$ resonance. The two mesons remain in a coherent
$p$-state until one of them decays.
The decay of one of the $B$ mesons into a flavor specific final state at time $t_{tag}$ projects the accompanying meson onto the opposite $b$-flavor at that time;
this meson decays to $\fcp$ at time $t_{CP}$.
The decay rates are now functions of $\delt\equiv t_{CP}-t_{tag}$ as
\begin{eqnarray}
\Gamma_{\bz(\bzb)\to\fcp}(\delt)=\af^2e^{-\Gamma |\delt|}\{1\pm\xi_f\sin2\phi_1\sin(\dmd \delt)\}.
\end{eqnarray}
Since $\delt$ is a signed value and integrated rates
do not have any asymmetry between $\bz$ and $\bzb$ mesons,
time dependent decay rates have to be measured to obtain $\sin2\phi_1$.

At KEKB, the $\Upsilon(4S)$ is produced with a Lorentz boost of $\beta\gamma=0.425$ along the electron beam direction ($z$ direction).
Since the $\bz$ and $\bzb$ mesons are nearly at rest in
the $\Upsilon(4S)$ center of mass system (cms), $\delt$ can be determined from the $z$ distance between the $f_{CP}$ and $f_{tag}$ decay vertices, $\Delta z\equiv z_{CP} - z_{tag}$, as $\Delta t \simeq \Delta z/\beta\gamma c$.
Precise measurement of the $B$ decay positions and proper
identification of the flavor of the accompanying $B$ meson are
crucial for the measurement of $\sin2\phi_1$.
In this paper, a determination of $\sin2\phi_1$ is presented along
with measurements of the $\bzd$-$\bzdb$ mixing parameter $\dmd$
and $B$-meson lifetimes.
The data sample corresponds to an integrated luminosity of $ 10.5~{\rm fb}^{-1}$
collected with the Belle detector~\cite{Belle} at the KEKB asymmetric $e^+e^-$ (3.5 on 8~GeV)  collider.~\cite{KEKB}

\begin{figure}
\begin{center}
\epsfxsize 2.8 truein \epsfbox{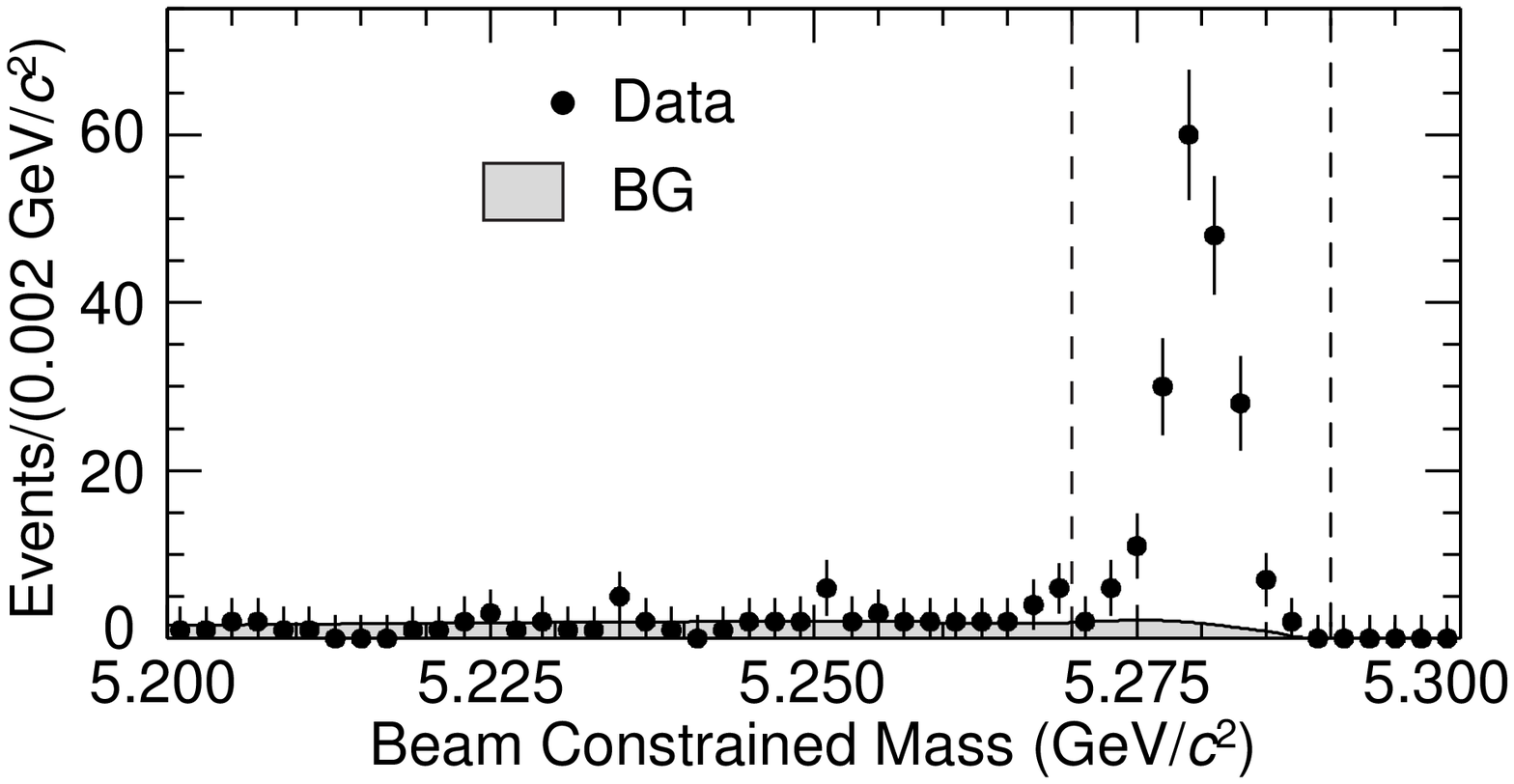}
\epsfxsize 2.57 truein \epsfbox{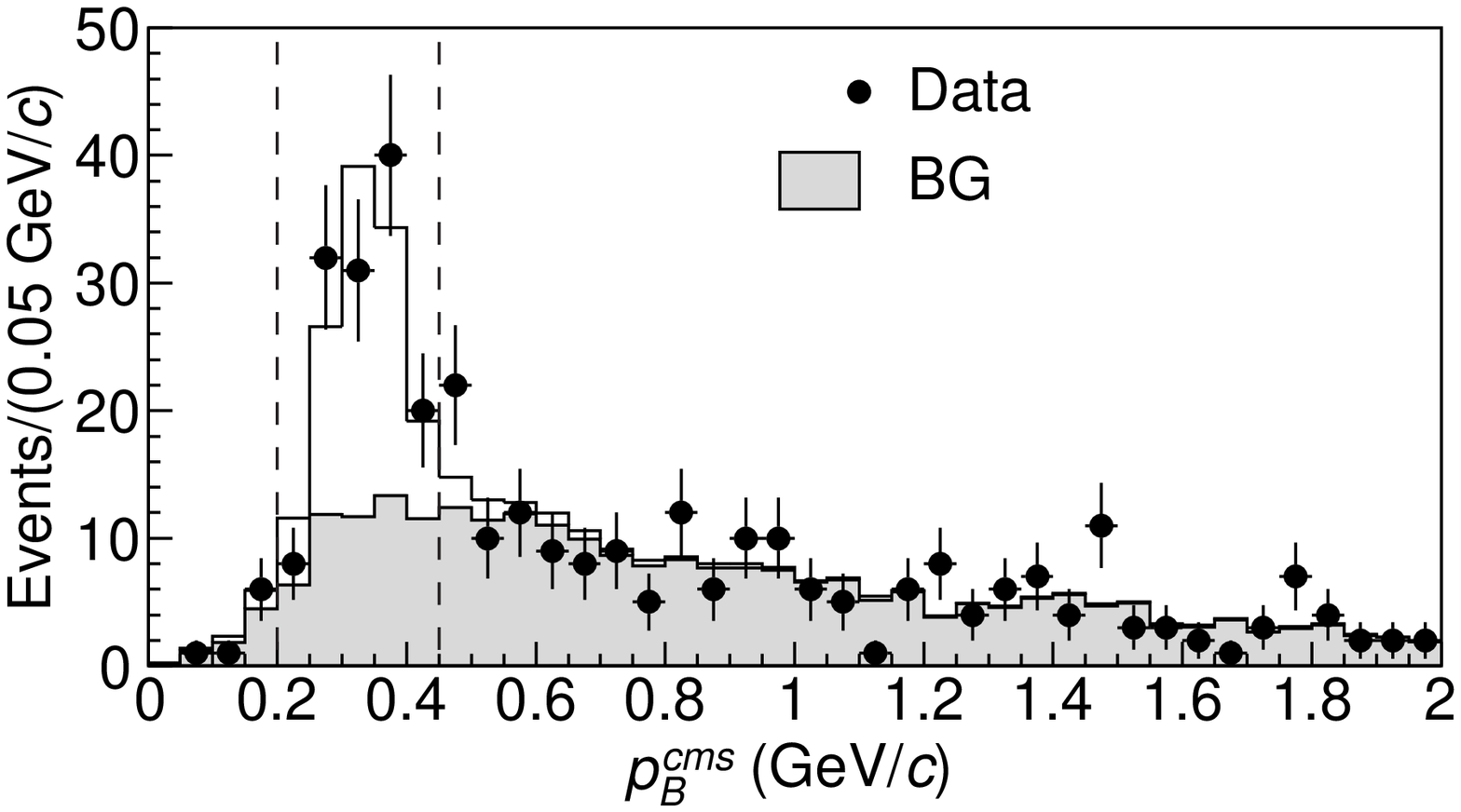}
\end{center}
\caption{(a) The beam-constrained mass distribution for
all $f_{CP}$ decay modes combined (other than  $B^0_d\to J/\psi \kl$).
The shaded area is the estimated background.
The dashed lines indicate the signal region.
(b) The $p_B^{cms}$ distribution for $B^0_d\to J/\psi \kl$ candidates
with the results of the fit.
The solid line is the signal plus background;
the shaded area is background only.
The dashed lines indicate the signal region.
}
\label{fig:bmass}
\end{figure}

\bigskip
The $\bz$ decay modes $\bz\to J/\psi\ks$, $\psi(2S)\ks$, $\chi_{c1}\ks$, $\eta_{c}\ks$, $J/\psi\kl$ and $J/\psi\piz$ are reconstructed for the $\sin2\phi_1$ measurement.
Fully reconstructed $B$ candidates are selected using
the energy difference $\Delta E\equiv E_B^{cms} - E_{beam}^{cms}$
and the beam-energy constrained
mass $M_{bc}\equiv\sqrt{(E_{beam}^{cms})^2-(p_B^{cms})^2}$,
where $E_{beam}^{cms}$ is the cms beam energy,
and $E_B^{cms}$ and $p_B^{cms}$ are the cms energy and momentum
of the $B$ candidate.
$p_B^{cms}$ is employed to select $J/\psi\kl$ events where $p_B^{cms}$ is calculated by a 0C fit using the $J/\psi$ momentum and the $\kl$ direction.
We reduce the background by means of
a likelihood quantity that
combines the $J/\psi$ cms momentum,
the angle between the $\kl$ and its nearest-neighbor charged track,
the charged track multiplicity, and
the kinematics that obtain when the event is reconstructed
assuming a $B^+ \to$ $J/\psi K^{*+},\;K^{*+}\to\kl\pi^+$ hypothesis.
Fig.~\ref{fig:bmass} shows (a) the $M_{bc}$ distribution for
all $f_{CP}$ decay modes combined (other than  $B^0_d\to J/\psi\kl$) and (b) the $p_B^{cms}$ distribution for $B^0_d\to J/\psi\kl$ candidates with the results of the fit.
We find 194 events and 131 events with expected backgrounds of 11 events and 54 events for all $\fcp$ modes without $\kl$ and $J/\psi\kl$ modes, respectively.

The flavor of the accompanying $B$ meson is identified using tracks such as;
high momentum leptons from $b\to c\ell^-\overline{\nu}$,
lower momentum leptons from $c\to s\ell^+\nu$,
charged kaons from $b\to c\to s$,
high momentum pions from decays of the type
$B_d^0\to D^{(*)-}(\pi^+, \rho^+, a_1^+, {\rm etc.})$, and
slow pions from $D^{*-}\to \overline{D}^0\pi^-$.
We calculate a value $Q_{track}=\{{\cal L}(\bz)-{\cal L}(\bzb)\}/\{{\cal L}(\bz)+{\cal L}(\bzb)\}$ in three track categories ($Q_\ell$ for lepton, $Q_K$ for kaon and $Q_\pi$ for pion) for tracks that are not associated with $f_{CP}$, where ${\cal L}$ is a likelihood based variables such as the particle identification information and the track momentum.
The best $Q$ values from three track categories are combined into one $Q$ value for each event using a three dimensional binned function $Q_{event}=f(Q_\ell, Q_K, Q_\pi)$ derived from a large statistics MC sample to take into account correlations between track categories.
In an ideal case, $|Q|$ should be related to the probabilities for an incorrect flavor assignment, $w$, as $|Q|=1-2w$.
The $w$ values are measured using $B^0_d\to D^{*-}\ell^+\nu$, $D^{(*)-}(\pi^+,\rho^+)$ modes.\cite{CC}
The $b$-flavor of the accompanying $B$ meson
is assigned according to the above-described flavor-tagging algorithm,
and values of $w$ are determined for six $|Q|$ intervals from the amplitudes of the
time-dependent $\bzd$-$\bzdb$ mixing oscillations.
We obtain $w_1=0.470^{+0.031}_{-0.035}\;(0<|Q|\le 0.25)$, $w_2=0.336^{+0.039}_{-0.042}\;(0.25<|Q|\le 0.5)$, $w_3=0.286^{+0.037}_{-0.035}\;(0.5<|Q|\le 0.625)$, $w_4=0.210^{+0.033}_{-0.031}\;(0.625<|Q|\le 0.75)$, $w_5=0.098^{+0.028}_{-0.026}\;(0.75<|Q|\le 0.875)$, $w_6=0.020^{+0.023}_{-0.019}\;(0.875<|Q|\le 1)$.
The total effective tagging efficiency is
$\sum_l f_l(1-2w_l)^2 = 0.270^{+0.021}_{-0.022}$,
where $f_l$ is the fraction of events in each $|Q|$ interval and the error includes both statistical and systematic uncertainties,
in good agreement with the MC result of 0.274.
We check for a possible bias in the flavor tagging
by measuring the effective tagging efficiency
for $B_d^0$ and $\overline{B_d}^0$ self-tagged samples separately,
and for different $\Delta t$ intervals.
We find no statistically significant difference.

The vertex positions for the $f_{CP}$ and $f_{tag}$ decays are
reconstructed using tracks
that have at least one
3-dimensional coordinate determined from associated $r\phi$ and $z$
hits in the same SVD layer
plus one or more additional $z$ hits in other SVD layers.
An interaction point constraint is applied to the vertex fit
for both $B$ mesons in order to improve the vertex resolution.
The average resolution estimated from data is 88~\micron\ for $z_{CP}$ and 164~\micron\ for $z_{tag}$.

We determine $\sin 2\phi_1$ from an
unbinned maximum-likelihood fit to the observed $\Delta t$ distributions.
The likelihood function is defined as
\begin{eqnarray*}
{\cal L}(\sin 2\phi_1)&=&\prod_i\int_{-\infty}^{\infty}d(\dtp) \{p_{SIG}^i(\dtp)R^i_{SIG}(\delt_i-\dtp)+p_{BG}^i(\dtp)R^i_{BG}(\delt_i-\dtp)\},\\
p_{SIG}^i(\dtp)&=&\frac{f_{SIG}^i}{2\cdot\taubz}e^{-\frac{|\dtp|}{\taubz}}\{1-q\xi_f(1-2w_l)\sin2\phi_1\sin(\dmd \dtp)\},\\
p_{BG}^i(\dtp)&=&\sum_k f_k^i \{(1-\flmbg^k)\cdot\delta(\dtp)+\flmbg^k\frac{\lmbg^k}{2}e^{-\lmbg^k|\dtp|}\},\\
\end{eqnarray*}
where: $q$ is the sign of the flavor tag variable $Q$; $R^i_{SIG}$ and $R^i_{BG}$ are the resolution functions calculated event-by-event from the track error matrix;
$f_{SIG}^i$ and $f_k^i$ are the fractions
of the signals and background contributions that are calculated event-by-event using
variables such as $\Delta E$, $M_{bc}$ and $p_B^{cms}$;
$\lmbg^k$, $\flmbg^k$ are the background-shape parameters.
A description of the $B$ reconstruction, vertex reconstruction, flavor tagging and fit procedure can be found elsewhere.\cite{tajima,Belle-CP}

\begin{figure}
\begin{center}
\epsfxsize 2.8 truein \epsfbox{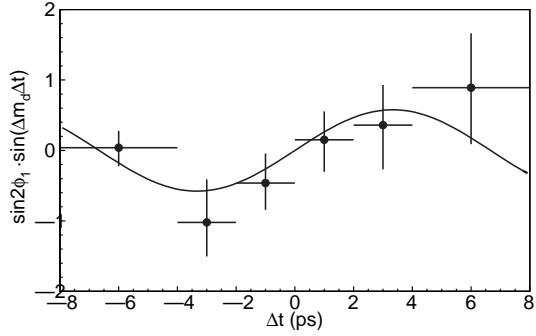}
\end{center}
\caption{ The asymmetry obtained
from separate fits to each $\Delta t$ bin;
the curve is the result of the global fit ($\sin 2\phi_1=0.58$).
}
\label{fig:cpfit}
\end{figure}

In the CP analysis, we obtain the result $\sin2\phi_1=0.58^{+0.32+0.09}_{-0.34-0.10}$,
where the first error is statistical
and the second systematic.
The systematic errors are  dominated by
the uncertainties in  $w_l$ ($^{+0.05}_{-0.07}$) and
the $J/\psi \kl$ background ($\pm 0.05$).
Separate fits to the $\xi_f=-1$ and  $\xi_f = +1$ event samples
give  $0.82^{+0.36}_{-0.41}$ and $0.10^{+0.57}_{-0.60}$, respectively.
Figure~\ref{fig:cpfit} shows the asymmetry obtained by performing the fit to events in  $\Delta t$ bins separately,
together with a curve that represents $\sin 2\phi_1\sin (\Delta m_d\Delta t)$
for $\sin 2\phi_1=0.58$.
We check for a possible fit bias by applying the same fit to  non-$CP$
eigenstate modes: $B^0_d \rightarrow D^{(*)-}\pi^+$, $D^{*-}\rho^+$,
$J/\psi K^{*0}(K^+\pi^-)$, and $D^{*-}\ell^+\nu$,
where ``$\sin 2\phi_1$'' should be zero, and the charged mode $B^+ \to J/\psi K^+$.
For all the modes combined we find $0.065\pm 0.075$, consistent with a null asymmetry.

In the $B$ lifetime analysis, $p_{SIG}(\dtp)$ is replaced by $p_{SIG}(\dtp)=\frac{f_{SIG}^i}{2\cdot\tau_B}e^{-{|\dtp|}/{\tau_B}}$.
The decay modes $\bz\to \dstm\ell^+\nu$ and $\bz\to D^{(*)-}(\pip, \rhop)$ are used for the $\bz$ lifetime measurements.
We obtain $\taubz=1.517\pm 0.045$~ps for the semileptonic mode and $\taubz=1.585^{+0.053}_{-0.051}$~ps for the hadronic modes, where the errors are statistical only. Combining these results, we obtain $\taubz=1.548^{+0.035}_{-0.034}$~ps.
Decay modes $\bplus\to \dstz\ell^+\nu$ and $\dz\pip$ are used for
the $\bplus$ lifetime measurement and give $\taubp=1.628\pm
0.060$~ps and $\taubp=1.679^{+0.049}_{-0.048}$~ps, respectively. The
combined result is $\taubp=1.656\pm 0.038$~ps.
Parameters for the resolution function $R_{SIG}$ are also determined in the lifetime fit and shared by the mixing and CP analysis.

In the mixing analysis, $p_{SIG}(\dtp)$ is replaced by $p_{SIG}(\dtp)=\frac{f_{SIG}^i}{4\cdot\taubz}e^{-{|\dtp|}/{\taubz}}\{1\pm(1-2w_l)\cos(\dmd\dtp)\}$ for unmixed ($\bzd\bzdb$) and mixed ($\bzd\bzd$ and $\bzdb\bzdb$) events.
Using $\bz\to\dstm\ell^+\nu$ and $\bz\to D^{(*)-}(\pip, \rhop)$ decay modes, we obtain $\dmd=0.522\pm 0.026$~ps${}^{-1}$ and $\dmd=0.527\pm 0.032$~ps${}^{-1}$, respectively.
Our results for the $B$ lifetime and the $\bzd$-$\bzdb$ mixing oscillation frequency $\dmd$ are consistent with the corresponding world averages.\cite{PDG}

\bigskip
We have measured the $B$ $CP$-violation parameter $\sin2\phi_1$, the
$\bzd$-$\bzdb$ mixing parameter $\dmd$ and $B$-meson lifetimes using 10.5 $fb^{-1}$ of data collected with the Belle detector at KEKB.
Neutral and charged $B$-meson lifetimes are measured to be $\taubz=1.548\pm0.035\;(\mathrm{stat.})$~ps and $\taubm=1.656\pm0.038\;(\mathrm{stat.})$~ps.
The oscillation frequency $\dmd$ for $\bzd$-$\bzdb$ mixing is measured to be $0.522\pm0.026\;(\mathrm{stat.})$~ps$^{-1}$ using semileptonic modes and $0.527\pm0.032\;(\mathrm{stat.})$~ps$^{-1}$ using hadronic modes.
The above results are preliminary.
The $B$ $CP$-violation parameter $\sin2\phi_1$ is measured to be
$$\sin 2\phi_1=0.58^{+0.32}_{-0.34}\;(\mathrm{stat.})^{+0.09}_{-0.10}\;(\mathrm{syst.}).$$
If the true value is 0, the probability to obtain $\sin2\phi_1> 0.58$ is 4.9\%.



\begin{thebibliography}{99}
\bibitem{CC}
Charge conjugate modes are implied throughout this paper.
\end{thebibliography}
\end{document}